\documentclass{aa}  

\usepackage{graphicx}
\usepackage{txfonts}
\begin{document} 

   \title{The focal-plane assisted pyramid wavefront sensor: enabling frame-by-frame optical gains tracking}

   \author{V. Chambouleyron
          \inst{1}\fnmsep\inst{2}
          \and
          O. Fauvarque\inst{3}
          \and
          J-F. Sauvage\inst{2}\fnmsep\inst{1}
          \and
          B. Neichel\inst{1}
          \and
          T. Fusco\inst{2}\fnmsep\inst{1}
          }

   \institute{Aix Marseille Univ, CNRS, CNES, LAM, Marseille, France\\
              \email{vincent.chambouleyron@lam.fr}
         \and
             DOTA, ONERA, Université Paris Saclay, F-91123 Palaiseau, France
          \and
          IFREMER, Laboratoire Detection, Capteurs et Mesures (LDCM), Centre Bretagne, ZI de la Pointe du Diable, CS 10070, 29280, Plouzane, France
             }

 
  \abstract
  {} 
  {With its high sensitivity, the Pyramid wavefront sensor (PyWFS) is becoming an advantageous sensor for astronomical adaptive optics (AO) systems. However, this sensor exhibits significant non-linear behaviours leading to challenging AO control issues.} 
  {In order to mitigate these effects, we propose to 
use, in addition to the classical pyramid sensor,  a focal plane image combined with a convolutive description of the sensor to perform a fast tracking of the PyWFS non-linearities, the so-called optical gains (OG).}
  {We show that this additional focal plane imaging path only requires a small fraction of the total flux, while representing a robust solution to estimate the PyWFS OG. Finally, we demonstrate the gain brought by our method with the specific examples of bootstrap and Non-Common Path Aberrations (NCPA) handling.}
   {}


   \maketitle
%

\section{Introduction}

The PyWFS, proposed for the first time in 1996 by \cite{raga}, is an optical device used to perform wavefront sensing. Inspired by the Foucault knife test, the PyWFS is a pupil plane wavefront sensor performing optical Fourier filtering thanks to a 4 faces glass pyramid located at the focal plane. The purpose of this glass pyramid is to split the electromagnetic (EM) field in four beams producing four different filtered images of the entrance pupil. This filtering operation allows the conversion of phase information at the entrance pupil into amplitude at a pupil plane where a quadratic sensor is used to record the signal (\cite{verinaud}, \cite{Guyon_2005}). Recently, the PyWFS has gained the interest of the astronomical community, as it offers a higher sensitivity than the classical Shack-Hartmann WFS commonly used in Adaptive Optics (AO) systems (\cite{esp}). However, the PyWFS exhibits non-linearities which prevent from having a simple relationship between the incoming phase and the measurements, leading to control issues in the AO loop. Previous studies (\cite{4k}, \cite{vdeo}) have demonstrated that one of the most striking impact of this undesirable behaviour is a time-averaged frequency-dependent loss of sensitivity when the PyWFS is working in presence of non-zero phase. This detrimental effect can be mitigated by providing an estimation of the so-called Optical Gains (OG): a set of scalar values encoding the loss of sensitivity with respect to each component of the modal basis. The goal of this paper is to present a novel way to measure the OG. In the first section we introduce the concept of Linear-Parameter Varying System (LPVS) to describe the PyWFS, consequently opening the possibility to estimate the OG frame by frame instead of considering a time-averaged quantity. In the second section, we present a practical implementation of the method, enabling the frame by frame OG tracking. Finally, we illustrate this OG tracking strategy in the context of closed-loop bootstrap and NCPA handling.

\section{The PyWFS seen as a Linear Parameter-Varying System (LPVS)}

\subsection{PyWFS non-linear behaviour and Optical Gains}

In the following, we call $s$ the output of the PyWFS. This output can be defined by different ways, the two main definitions are called "full frame" or "slopes maps". In the first case, $s$ is obtained by recording the full  image of the WFS camera, for which a reference image corresponding to a reference phase has been removed (\cite{FauvOptica}). In the second case, the WFS image is processed to reduce the useful information to two pupil maps usually called "slopes maps" (\cite{raga}). The work presented here remains valid for both full-frame or slopes-map computation, and we decided to use the full frame definition for all the rest of the paper. \\

When described with a linear model, the PyWFS output are linked with the incoming phase $\phi$ through an interaction matrix called $M$. This interaction matrix can be built thanks to a calibration process which consists in sending to the WFS a set of phase maps (usually a modal or a zonal basis) with the deformable mirror (DM) and then record the derivative $\delta s(\phi_{i})$ of the PyWFS response for each component of the basis. This operation is most commonly done through the so-called push-pull method, consisting in sending successively each mode with a positive and then negative amplitude $a$ to compute the slope of the linear response:

\begin{equation}
    \delta s(\phi_{i}) = \frac{s(a\phi_{i})-s(-a\phi_{i})}{2a}\
.\end{equation}

The interaction matrix (also called Jacobian matrix) is then the collection of the slopes recorded for all modes:

\begin{equation}
    M = (\delta s(\phi_{1}),..., \delta s(\phi_{i}),..., \delta s(\phi_{N}))
.\end{equation}

\begin{figure}[!h]
    \centering
    \includegraphics[scale=0.6]{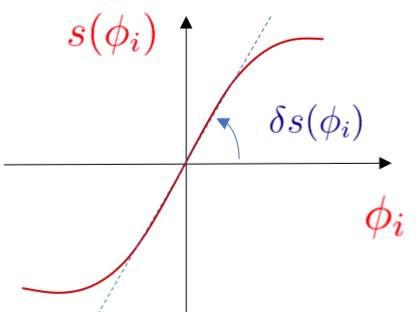}
    \caption{Sketch of the PyWFS response curve for a given mode $\phi_{i}$. The push-pull method around a null-phase consists in computing the slope of this curve at $\phi_{i}=0$.}
    \label{fig:linearityCurve}
\end{figure}

In that linear framework, we can then link the measured phase with the output of the PyWFS through the relationship:

\begin{equation}
    s(\phi)  = M.\phi
    \label{eq:linear_framework}
\end{equation}

This matrix computation formalism has interesting properties that are required in the AO control loop. However, the PyWFS exhibits substantial non-linearities which make the equation above only partially true. Mathematically, the deviation from linearity is expressed thanks to the following inequality: $s(a\phi_{i}+\phi)\neq s(a\phi_{i})+s(\phi)$, where $\phi$ is a non-null given phase. When working around $\phi$, the slope of the linear response of the sensor is therefore modified:

\begin{equation}
\begin{split}
    \delta s_{\phi}(\phi_{i}) &= \frac{s(a\phi_{i}+\phi)-s(-a\phi_{i}+\phi)}{2a}\\
    &\neq \delta s(\phi_{i})
    \label{eq:NonLinear}
\end{split}
\end{equation}

During AO observation, the sensor is working around a non-null phase $\phi$ corresponding to the residual phase of the system. As a consequence of Equation \ref{eq:NonLinear}, the response of the system will be modified. Previous studies suggest to update the response slopes to mitigate this effect by relying on two main concepts: 
\begin{itemize}
    \item The stationarity of the residual phases (\cite{rigaut98}). For a given system and fixed parameters (seeing, noise, etc...) one can compute an averaged response slope for each mode. It has been proven (\cite{fauv}) that under this stationarity hypothesis, the averaged response slope depends on the statistical residual phases behaviour through their structure function ($D_{\phi}$): $<\delta s_{\phi(\phi_{i})}> = \delta s_{D_{\phi}}(\phi_{i})$.
   \item The diagonal approximation (\cite{4k}). This approximation implies to consider no cross-talk between the modes, which means that the response slopes are only modified by a scalar value for each mode. This value is known as the Optical Gain (OG). We then have $\delta s_{D_{\phi}}(\phi_{i}) = t^{i}_{D_{\phi}}.\delta s(\phi_{i})$, where $t^{i}_{D_{\phi}}$ is the OG associated to the mode $i$ for a given residual phase perturbation statistics characterised by the structure function $D_{\phi}$. In this approximation, the shape of the response is left unchanged.
\end{itemize}

Finally, the update of interaction matrix is simply done by multiplying by a diagonal matrix $T_{D_{\phi}}$ called the OG matrix, whose diagonal components are $t^{i}_{D_{\phi}}$.  
\begin{equation}
\begin{split} 
    s(\phi) & = <M_{\phi}>.\phi\\
    & = M_{D_{\phi}}.\phi \\
    & \approx M.T_{D_{\phi}}.\phi
\end{split} 
\end{equation}

We use the scalar product presented in \cite{chambou} to calculate the diagonal components of this matrix:

\begin{equation}
     T_{D_{\phi}} =\frac{\text{diag}(M^{t}_{D_{\phi}}M)}{\text{diag}(M^{t}M)}
\end{equation}

Several approaches to practically compute this matrix can be found in the literature. They can be split in two categories: the ones which are invasive for the science path, consisting in sending some probe modes to the DM to get back to the OG (\cite{espositoNCPA}, \cite{vdeo}), and the ones which rely on the knowledge of the statistics of the residual phases through the telemetry data to estimate the OG (\cite{chambou}). In all the proposed methods, the OG can be seen as evaluation of a time-averaged loss of sensitivity of the sensor. Being able to accurately retrieve OG allows to compensate for the sensitivity loss.  

\subsection{The LPVS approach}

As described by equation \ref{eq:NonLinear}, the PyWFS outputs are impacted by the incoming phase. The time-averaged definition of the interaction matrix $M_{D_{\phi}}$, although having good properties, is limited to a statistical behaviour of the PyWFS. 
In this paper, we propose a framework that will address the non-linearites in real time, with an interaction matrix updated at every frame. To do so, we first assume that the diagonal hypothesis holds. Then, and inspired by the automatic field domain, the PyWFS is now considered as a Linear-Parameter Varying System (LPVS) (\cite{LPVS}): its linear behaviour encoded by the interaction matrix is modified at each frame according to the incoming phase. Under this framework, the new expression of the PyWFS output can be written as:
\begin{equation}
    s(\phi) = M_{\phi}.\phi \approx M.T_{\phi}.\phi
\end{equation}

where $T_{\phi}$ is the OG matrix for the given measured phase $\phi$. Assuming the diagonal approximation holds, one can extract $T_{\phi}$ from the interaction matrix computed around $\phi$:

\begin{equation}
     T_{\phi} =\frac{\text{diag}(M^{t}_{\phi}M)}{\text{diag}(M^{t}M)}
     \label{eq:OGdef}
\end{equation}

Note that for a given system, repeating this operation on a set of different phases will eventually lead to the time-averaged definition of the OG matrix:

\begin{equation}
     <T_{\phi}> =T_{D_{\phi}}
\end{equation}

To illustrate the difference between the time-averaged response and a single realisation, we performed simulation presented in Figure \ref{fig:fastOG}. These simulations are done with parameters consistent with an 8m telescope and for two seeing conditions. Note that all results showed in this paper rely on end-to-end simulations performed via the \textit{OOMAO} \textsc{Matlab} toolbox (\cite{oomao}). The exact conditions and parameters are summarised in Table \ref{tab:simu}. 
In simulation, one can compute the exact PyWFS response, by freezing the entrance phase and performing a calibration process around this working point. We therefore computed $T_{\phi}$ for 1000 residual phases realisations, and show the OG variability for two seeing conditions in Figure \ref{fig:fastOG}. Note that this represents an optimistic context where the Fried parameter $r0$ is fixed through the complete simulation. By estimating the $T_{\phi}$ with a time averaging strategy, the errors on the OG corresponding to a given residual phase can reach more than tens of percent (OG exhibiting a maximum deviation from the averaged value are enlightened). This result illustrates the potential gain of performing a frame-by-frame estimation of the OG instead of a time-averaged one. In the next section, a practical mean to perform this frame by frame gain scheduling operation is presented.\\

\bgroup
\def\arraystretch{1.5}
\begin{table}[!h]
\small
\begin{tabular}{ |p{2cm}|p{5cm}| }
\hline
\hline
Resolution & 80 pixels in telescope diameter  \\
\hline
Telescope & D = 8 m - no central obstruction  \\
 \hline
Atmosphere & Von-Karman PSD - 3 layers \\
 \hline
Deformable mirror  & Generating atmospheric Karhunen-Loève (KL) basis: 400 modes \\
 \hline
Sensing Path & $\lambda = 550\ nm$ - 40 subpupils in D \\
\hline
\hline
\end{tabular}
\vspace{1pt}
\caption{Simulation parameters.}
    \label{tab:simu}
\end{table}

\begin{figure}[!h]
    \centering
    \includegraphics[scale=0.45]{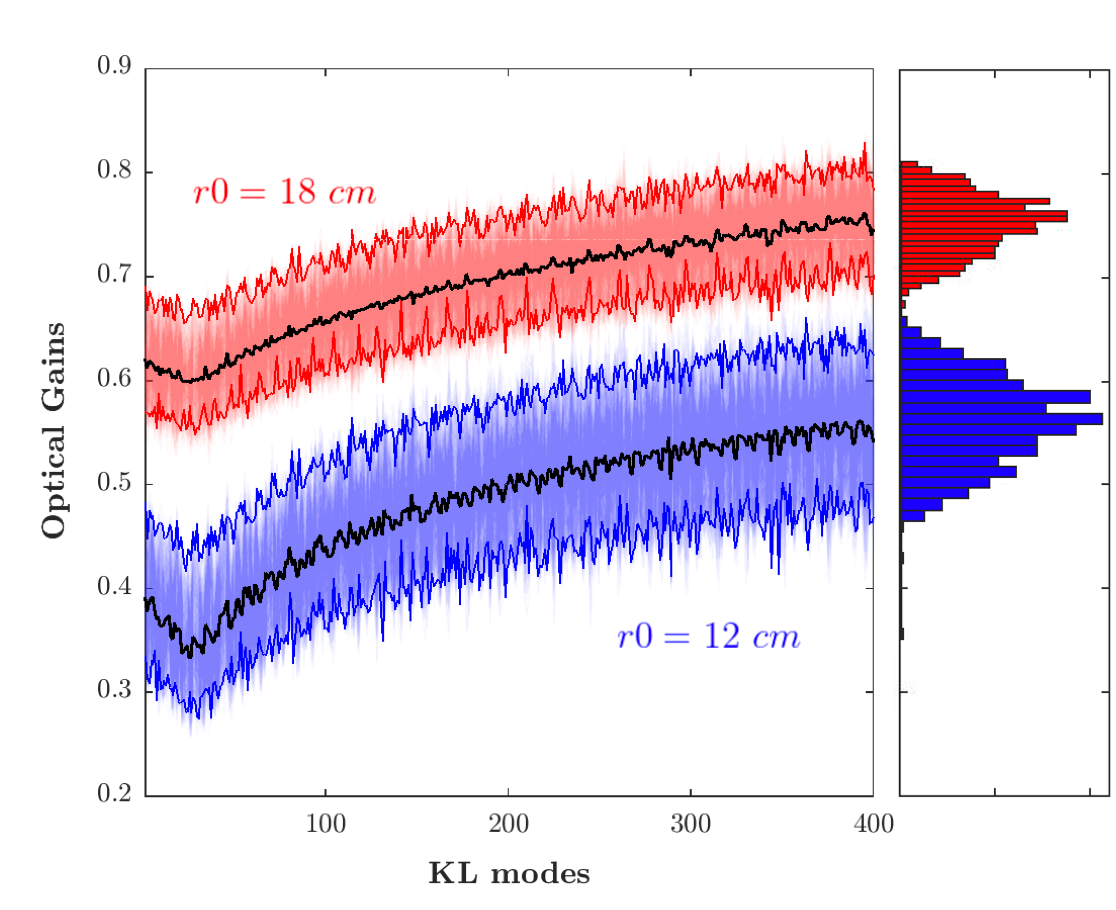}
    \caption{ Variability of closed-loop OG. For given system parameters we compute $T_{\phi}$ for $1\ 000$ phase realisations in two seeing configurations: $r0 = 18\ cm$ and $r0 = 12\ cm$. The variability of the frame by frame OG is exposed through the histogram on the right part of the Figure, and by the highlighted extreme OG curves for each $r0$ case.}
    \label{fig:fastOG}
\end{figure}

\section{Gain scheduling camera}

\subsection{Principle}

Getting an estimation of the OG values (the diagonal of $T_{\phi}$) requires to get an additional information describing the working point of the PyWFS at each moment, independently from the PyWFS measurements themselves. To this end, a specific sensor called Gain Scheduling Camera (GSC) is implemented.

Empirically, it is well known that the PyWFS sensitivity depends on the structure of the EM field when it reaches the pyramid mask. For instance, the more this field is spread over the pyramid mask, the less sensitive the PyWFS is. Besides, because sensitivity and dynamic range are antagonist properties, a well-known technique used to increase PyWFS dynamic range consists in modulating the EM field around the pyramid apex. In order to keep track of the sensor regime, we consequently suggest to probe this EM field by acquiring a focal plane image synchronously with the Pyramid WFS data. This can be simply done by placing a beam-splitter before the pyramid mask and recording the signal on a focal plane camera having the same field of view of the pyramid (Fig. \ref{fig:modCam}). 

\begin{figure}[!h]
    \centering
    \includegraphics[scale=0.75]{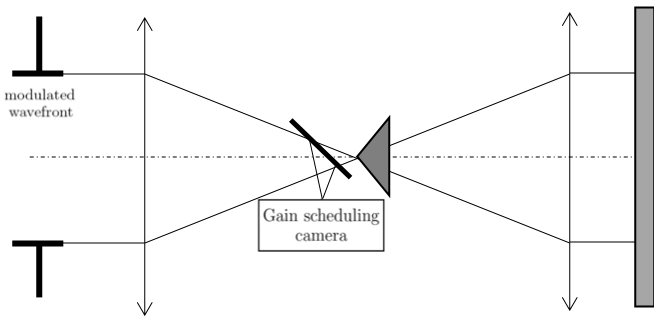}
    \caption{The gain scheduling camera is a focal plane camera which records the intensities of the modulated EM field with same pyramid field of view. This operation requires to use part of the flux from the pyramid path.}
    \label{fig:modCam}
\end{figure}

\begin{figure}[!h]
    \centering
    \includegraphics[scale=0.4]{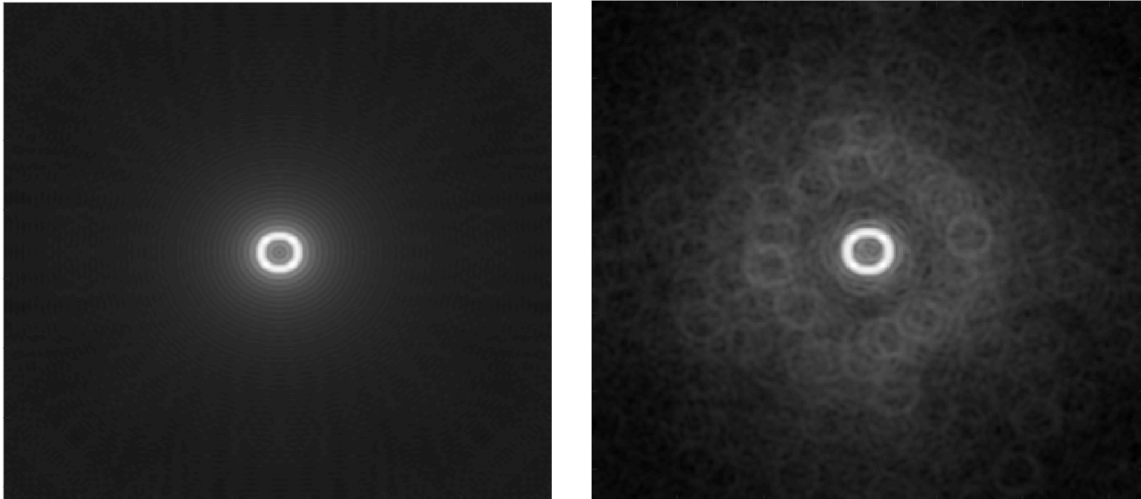}
    \caption{\textbf{Left:} Gain scheduling camera image for a flat wavefront. The white circle is produced by the Tip Tilt modulation of the Pyramid signal. \textbf{Right:} Gain scheduling camera image for a given closed-loop residual phase.}
    \label{fig:psfs}
\end{figure}

In that configuration, the focal plane camera - hereafter called the GSC - is recording the intensity of the modulated EM field seen by the pyramid. By using the same exposure time and frame rate as the WFS camera, the signal observed is then an instantaneous AO-corrected Point-Spread Function (PSF) convolved with the circle of modulation. This is illustrated by Figure \ref{fig:psfs}, where one can see the modulation circle on the left, and the replicas of this modulation circle by the focal plane speckles on the right. By denoting $\Omega_{\phi}$ the GSC signal, we can therefore write:

\begin{equation}
    \Omega_{\phi}=\text{PSF}_{\phi}\star \omega
\end{equation}

where $\omega$ is the modulation weighting function. This latter can be thought of as a map of the incoherent positions reached by the EM field on the pyramid during one integration time of the WFS camera. This function is thus a circle for the circularly modulated PyWFS (Figure \ref{fig:IRparam} - Right). $\Omega_{\phi}$ has to be understood as the effective modulation weighting function: the phase to be measured is producing its own modulation leading to PyWFS loss of sensitivity, and the GSC is therefore a way to keep track of this additional modulation.\\

The next step is now to link this focal plane information with the PyWFS optical gains and merge both GSC and PyWFS signal in one final set of WFS outputs. In a previous work (\cite{chambou}), we demonstrated that the convolutive model of the PyWFS developed by \cite{fauv} can be used to predict the averaged OG if the statistical behaviour of the residual phases (through the knowledge of their structure function) is known. In equation \ref{eq:IR} we remind the expression of the PyWFS output in this convolutive framework.

\begin{equation}
    s(\phi)= \textbf{IR}\star(\mathbb{I}_{p}\phi)
    \label{eq:IR}
\end{equation}

where $\textbf{IR}$ is the impulse response of the sensor and $\star$ the convolutive product. In the framework of the infinite pupil approximation, the impulse response around a flat wavefront can be expressed through two quantities, the mask complex function $m$ and the modulation function $\omega$ (Figure \ref{fig:IRparam}):

\begin{equation}
    \textbf{IR} = 2\text{Im}(\overline{\widehat{m}}(\widehat{m}\star\widehat{\omega}))
    \label{eq:IRbis}
\end{equation}

\begin{figure}[!h]
    \centering
    \includegraphics[scale=0.7]{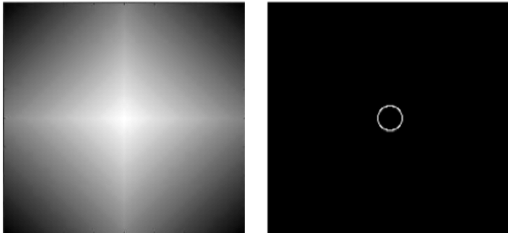}
    \caption{ \textbf{Left:} $arg(m)$ - shape of the pyramid phase mask in the focal plane. \textbf{Right:} $\omega$ - the modulation weighting function: different positions reached by the EM field during one integration time.}
    \label{fig:IRparam}
\end{figure}

We propose here to combine this model with the signal delivered by the GSC in order to compute the impulse response $\textbf{IR}_{\phi}$ of the PyWFS around each individual phase realisation. For that, we simply replace $\omega$ by the gain scheduling camera data as described by equation \ref{eq:IR_cam}. 

\begin{equation}
    \textbf{IR}_{\phi} = 2\text{Im}(\overline{\widehat{m}}(\widehat{m}\star\widehat{\Omega_\phi}))
    \label{eq:IR_cam}
\end{equation}

This new way to compute the impulse response can be seen as using the impulse response given for an infinite pupil system (Eq.\ref{eq:IRbis}) for which we replaced the modulation weighting function by the energy distribution at the focal plane, including both the modulation and the residual phase. \\
Now that we are able to compute $\textbf{IR}_{\phi}$ at each frame, we can get an estimation of the OG matrix $\widetilde{T}_{\phi}$ through the following computation of its diagonal components already described in \cite{chambou}: 

\begin{equation}
     \tilde{t}^{\ i}_{D_{\phi}}=\frac{\langle\textbf{IR}_{\phi}\star\phi_{i}|\textbf{IR}_{\text{calib}}\star\phi_{i}\rangle}{\langle\textbf{IR}_{\text{calib}}\star\phi_{i}| \textbf{IR}_{\text{calib}}\star\phi_{i}\rangle}
     \label{eq:Gconv}
\end{equation}

where $\textbf{IR}_{\text{calib}}$ is the impulse response computed for the calibration state, most commonly for $\phi = 0$ (Figure \ref{fig:psfs}, left). \\

\subsection{Accuracy of the estimation}

It is now possible to test the accuracy of our estimator by comparing $\widetilde{T}_{\phi}$ and $T_{\phi}$. For that, we compute the "true" $T_{\phi}$ through End-to-End simulations, by proceeding through the ideal way described in section above: an interaction matrix is computed around each given residual phase, from which the OG matrix is derived (Eq. \ref{eq:OGdef}). This will provide the "ground truth", to which the gains estimated with the GSC will be compared.\\
First results are shown in Figure \ref{fig:OG_cam}, under different seeing and modulation conditions. As illustrated by Figure \ref{fig:OG_cam}, there is a very nice agreement between the real and estimated OG, demonstrating the accuracy of the proposed method.


\begin{figure}[!h]
    \centering
    \includegraphics[scale=0.45]{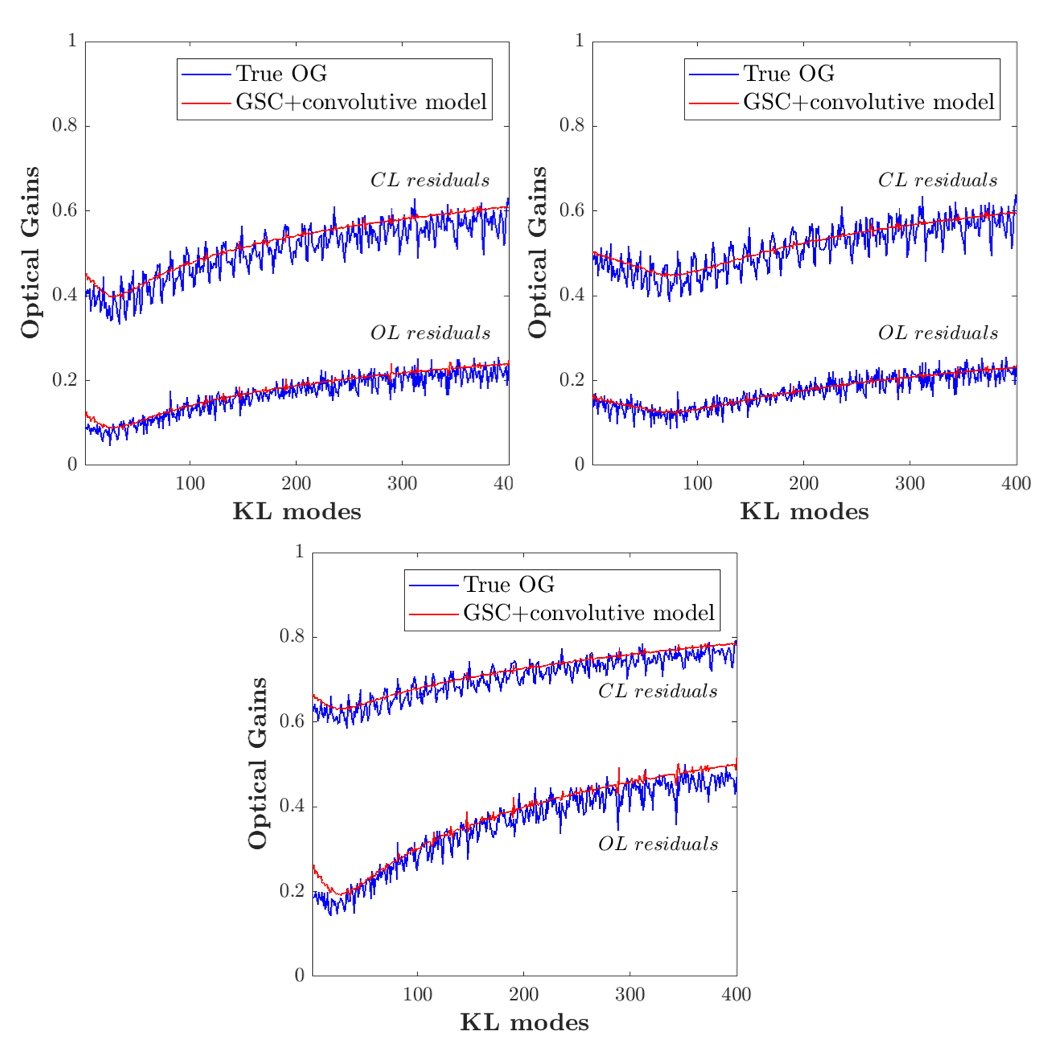}
    \caption{ OG estimation for given residual phases thanks to the GSC are compared with end-to-end simulation for different parameters (same framework as in Figure \ref{fig:fastOG}). OL: open-loop/CL: closed-loop residual phases \textbf{Left:} $r0=12\ cm$ and $r_{mod}=3\lambda/D $. \textbf{Middle:} $r0=12\ cm$ and $r_{mod}=5\lambda/D$. \textbf{Right:} $r0=18\ cm$ and $r_{mod}=3\lambda/D$.}
    \label{fig:OG_cam}
\end{figure}

For the parameters used in our simulations, the estimation remains accurate whether we are in open-loop or closed loop. The ripples seen in the "ground truth" OG curves are smoothed in the convolutive framework. As a matter of fact, the convolutive product given Equation \ref{eq:IR} tends to smooth the output of the PyWFS even when the impulse response is computed around a non-zero phase. On Figure \ref{fig:OG_cam}, we can also spot a slight deviation for low order modes when having a low modulation regime and for strong entrance phase (open-loop here).

\subsection{Robustness to noise}

The GSC has shown to be a reliable way to perform a fast OG tracking, but it requires to use a fraction of the photons available in the sensing path. This will inevitably compete with the gain of sensitivity provided by the PyWFS. The goal of this section is then to demonstrate that our GSC approach is only weakly impacted by photon noise, therefore requiring only a small amount of photons while performing an accurate frame by frame OG estimation. To this end, we propose to inject noise in the data delivered by the GSC and to probe the impact on the OG estimation. 

We run simulations with the same parameters described above. The sensing path is working around the central wavelength $\lambda_{c} = 550\ nm$ with the given bandwidth $\Delta\lambda = 90\ nm$ and an ideal transmission of 100 percent. 
The exposure time of the GSC is 2 milliseconds (frame rate of the loop) and 10 percent of the photons are used by the GSC camera. The GSC pixel size corresponds to Shannon sampling of the diffraction-limited PSF. In this given configuration, the data recorded by the GSC for a given closed-loop residual phase ($r0 = 14\ cm$, $r_{mod}=3\ \lambda/D$) is presented in Figure \ref{fig:magnitude} (Top part) for: \textit{(a.)} A noise-free system \textit{(b.)} A guide star magnitude equal to 8, \textit{(c.)} A guide star magnitude equal to 10 and \textit{(d.)} A guide star magnitude equal to 12. For these three noise configurations ($mag = 8, 10$ and $12$) we estimate the OG for 500 realisations of the noise. The results are given Figure \ref{fig:magnitude} (Bottom part). It can be seen that the introduction of noise leads to an increased OG estimation error, which logically scales with the signal-to-noise ratio accordingly to $\sqrt{n_{ph}}$. However, it can also be seen that the GSC approach still performs a satisfactory OG estimation even for low-magnitude guide stars. Note also that in the case when dealing with even fainter guide stars, one could mitigate the noise impact by integrating the GSC data over several frames. A trade-off between noise propagation and OG error would then be required.\\

\begin{figure}[!h]
    \centering
    \includegraphics[scale=0.45]{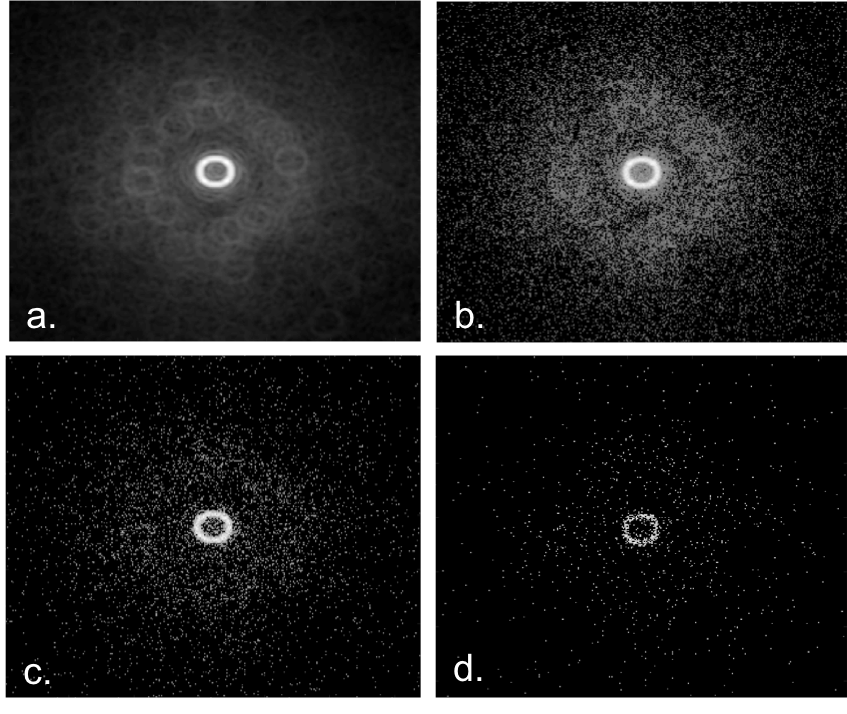}
    \includegraphics[scale=0.45]{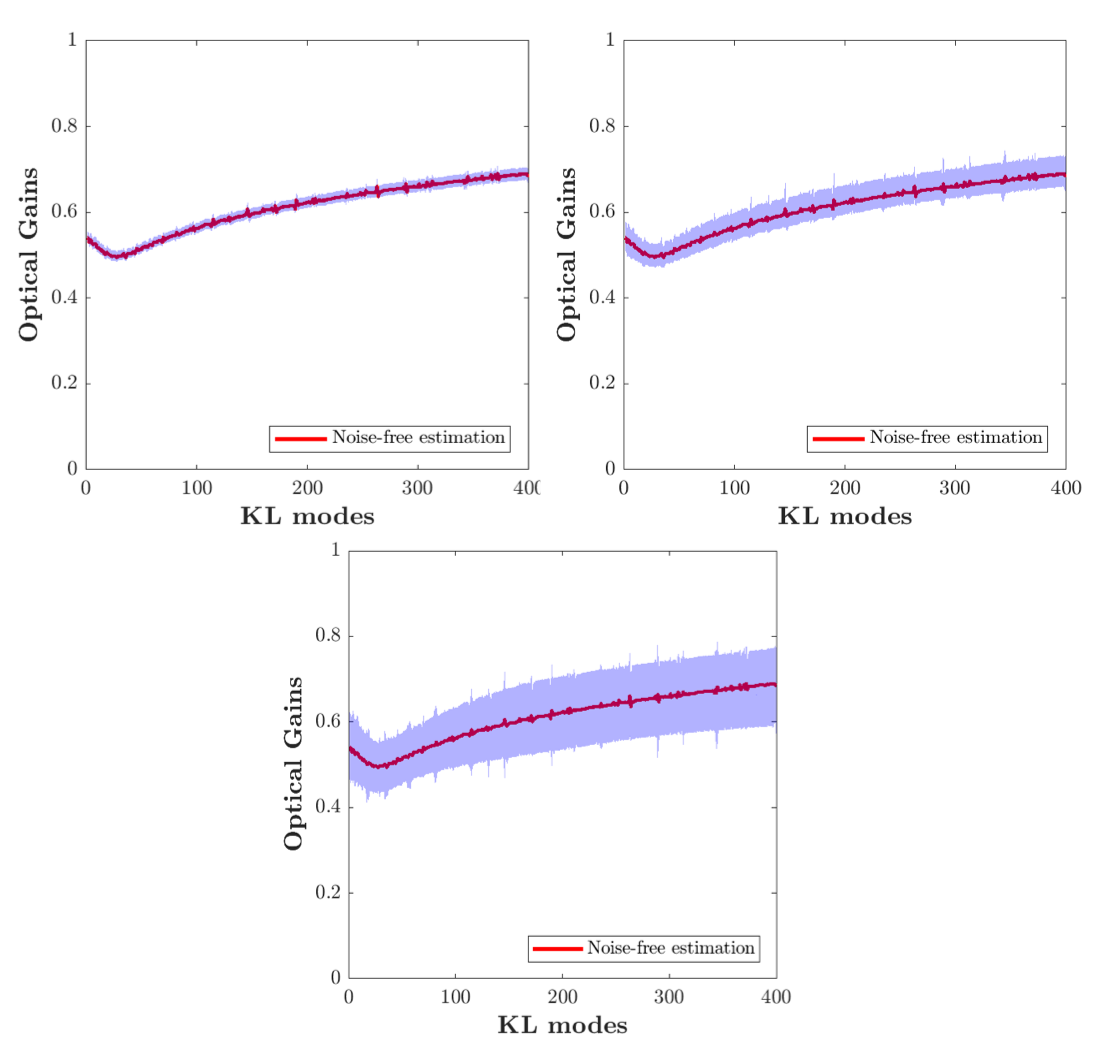}
    \caption{\textbf{Top:} Closed-loop GSC images for different entrance fluxes. In the chosen configuration, the exposure time is 2 ms and we collect 10\% of the photons in the sensing path. \textbf{a:} Infinite number of photons. \textbf{b:} Guide star magnitude = 8 ($n_{ph} = 55\ 000$ on the GSC). \textbf{c:} Guide star magnitude = 10 ($n_{ph} = 9\ 000$ on the GSC). \textbf{d:} Guide star magnitude = 12 ($n_{ph} = 1\ 400$ on the GSC). \textbf{Bottom:} OG estimation for the noise-free system compared with the three noisy configurations.}
    \label{fig:magnitude}
\end{figure}

These results are crucial because they demonstrate that one can use the GSC with only a small fraction of WFS photons, leading to a limited repercussion on the signal-to-noise ratio on the PyWFS. We therefore have a way to estimate the OG , and to some extend increase the linearity of the sensor, while having a reduced impact on its sensitivity.

\subsection{GSC spatial sampling}

Another aspect is the sampling of the GSC detector with respect to the modulated PSF. Indeed, if an under-sampling could be considered, it would reduce the number of pixels required by the GSC, and consequently reduce the practical implementation complexity. To test this, we ran our algorithm for various samplings of the GSC, in order to see the impact on the OG estimation. The results for a given closed-loop residual phase ($r0 = 14\ cm$, $r_{mod}=3$) are given in  Figure \ref{fig:shannonOG}. One can see that the sampling of the PSF can go below the Shannon sampling (2 pixels per $\lambda/D$) without significant impact on the estimation. In fact, this result depends on the modulation radius $r_{mod}$ used and we noticed that the OG estimation is not affected as long as the pixel size $d_{px}$ satisfies the Shannon criterion for the modulation radius, \textit{i.e}: 

\begin{equation}
     d_{px} \le r_{mod}/2
     \label{eq:px_size}
\end{equation}

As soon as this criterion is not respected, the under sampled modulation circle is seen as a disc (Figure \ref{fig:shannonOG}), impacting the OG estimation for low order modes. \\
As a concrete example, a PyWFS for the Extremely Large Telescope (ELT) working at $\lambda = 800\ nm$ with a field of view of 2 arcsecs and with a sampling of Shannon/4 on the GSC would require a GSC camera with no more than $250\times 250\ px$. This limited size allows for the use of low-readout noise cameras as GSC, and remaining in a photon-noise limited regime.  \\

\begin{figure}[!h]
    \centering
    \includegraphics[scale=0.55]{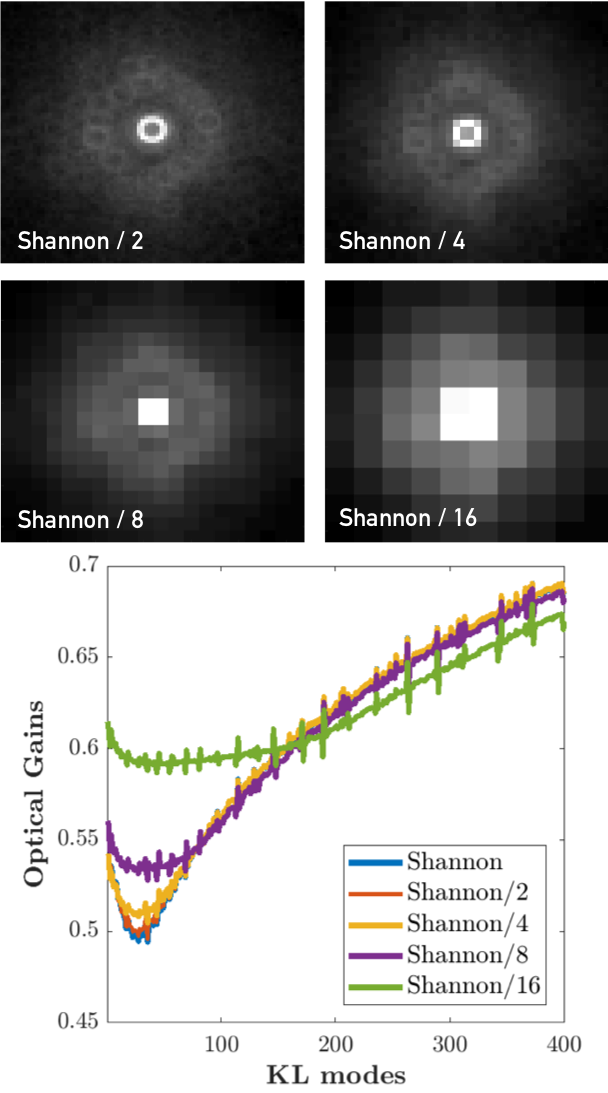}
    \caption{Impact of the GSC sampling on OG estimation for a given closed-loop residual phase ($r0 = 14\ cm$, $r_{mod}=3$). \textbf{Top:} Images delivered by the GSC with different samplings.  \textbf{Bottom:} Impact on the OG estimation.}
    \label{fig:shannonOG}
\end{figure}

To conclude this section, we have shown that it was possible to perform OG fast-tracking by using an image of the modulated EM field at the focal plane. Our method uses a so-called Gain Scheduling Camera providing a non-biased information on the working point of the PyWFS, and the subsequent OG estimation using a convolutive model. We demonstrated that the GSC can work with a limited number of photons and pixels, which makes the practical implementation fully feasible. The next section is dedicated to quantifying the performance benefits of OG fast tracking with the GSC.

\section{Application to specific AO control issues: bootstrapping and NCPA handling}

As shown in the previous sections, the GSC allows one to track frame-by-frame the PyWFS OG, and compensate for these non-linearities. We illustrate here two possible situations where the GSC can significantly improve the performance: bootstrapping and NCPA handling.

\subsection{Bootstrapping}

During the AO loop bootstrap, the PyWFS is facing large amplitude wavefronts (due to uncorrected turbulence) leading to significant non-linearities that may prevent the loop from closing. Therefore, this step is critical because it corresponds to the moment where the OG are the most important. Keeping track of them frame by frame in order to update the reconstructor helps closing the AO loop. Given the timescales involved in the AO loop bootstrap, this problem cannot be tackled by other OG handling techniques previously studied in the literature. The best solutions already proposed endures necessarily delays of few frames (\cite{close}). Here, we can estimate the OG corresponding to the current measurement frame: it is an unprecedented feature. We show in Figure \ref{fig:OG_bootstrap} different images delivered by the GSC during the bootstrap operation. The corresponding estimated OG are also plotted, compared with the End-to-End computation giving the true OG values. While the loop is closing, the OG varies from low values to higher values indicating that the residual phases reaching the PyWFS are lowering: the loop is closing and the DM is starting to correct the atmospheric aberrations. Our technique manages to do a precise OG follow-up during all the steps of the process, at the frame of the loop.\\

\begin{figure}[!h]
    \centering
    \includegraphics[scale=0.55]{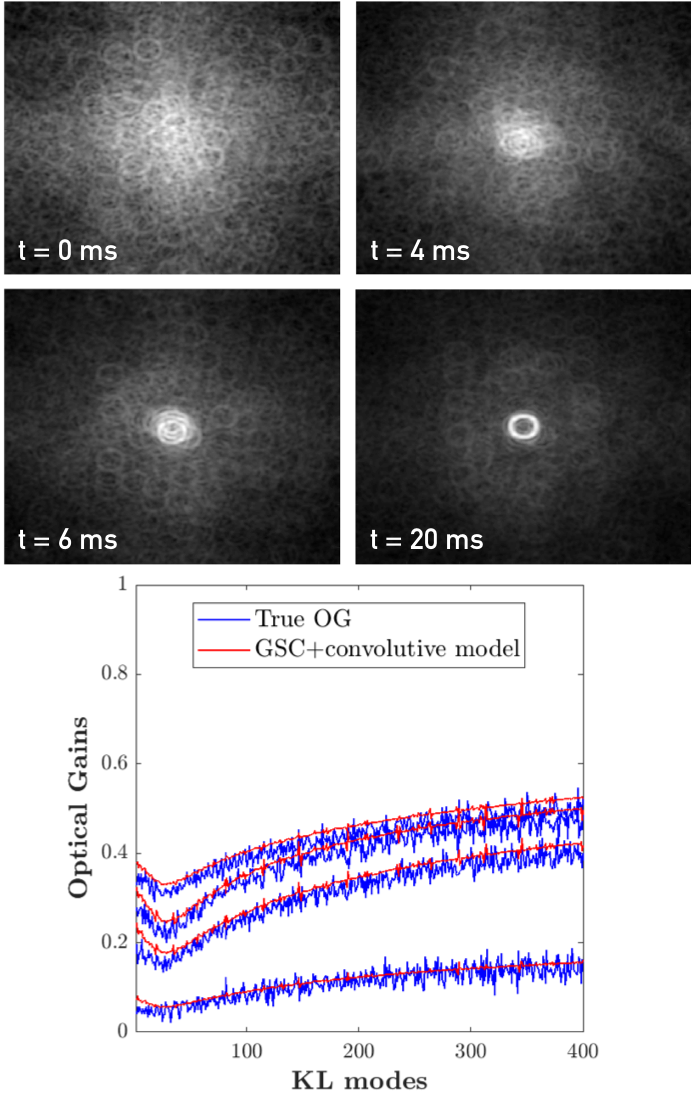}
    \caption{ Bootstrapping with the help of the GSC. \textbf{Top:} Images delivered by the GSC at time. $t=0$ represents the beginning of the servo-loop. The framebrate of the AO loop is still fixed at 2 ms with $r0=12\ cm$ and $r_{mod}=3\lambda/D$. \textbf{Bottom:} OG estimation during bootstrap, for the corresponding images on the left. Lower OG correspond to higher residuals on the pyramid, hence to the first frames of loop closure.}
    \label{fig:OG_bootstrap}
\end{figure}

We can use our frame by frame OG estimation to update the reconstructor while the loop is closing. The reconstructor is the pseudo-inverse of the interaction matrix, we can therefore relate it to the OG matrix and the calibration interaction matrix through the following formula: 

\begin{equation}
        M_{\phi}^{\dagger} = T_{\phi}^{-1}M_{\text{calib}}^{\dagger}        
    \label{eq:update_Recon}
\end{equation}

By doing so, we show that it is possible to close the loop faster. A simulation example is presented through the comparison of a loop bootstrap without OG compensation and with OG compensation thanks to the GSC camera (Figure \ref{fig:SR_bootstrap}). This example, with a limited benefit in practice, shows how a fast OG tracking combined with the corresponding update of the reconstructor can be applied to mitigate all kind of short timescale residuals variations, like seeing bursts for instance.

\begin{figure}[!h]
    \centering
    \includegraphics[scale=0.4]{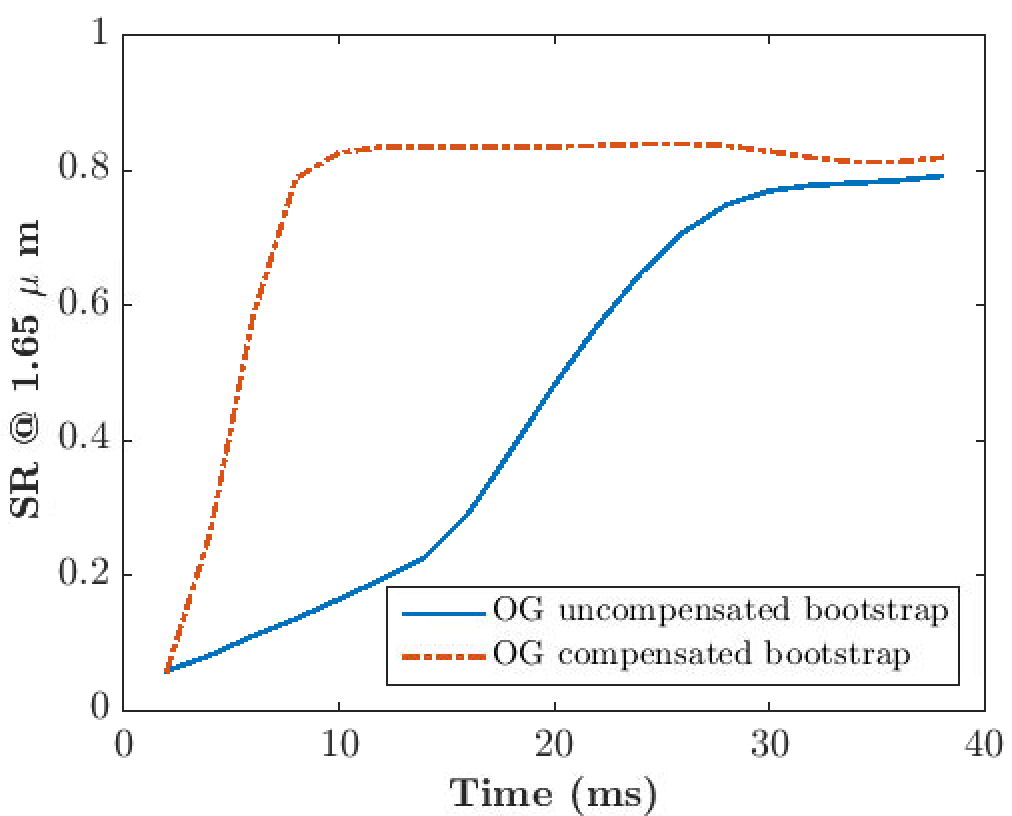}
    \caption{OG compensated bootstrap versus OG uncompensated bootstrap.}
    \label{fig:SR_bootstrap}
\end{figure}

\subsection{NCPA handling}

NCPA handling is emerging as one of the main issues due to PyWFS OG, as it was for instance demonstrated on the Large-Binocular-Telescope (\cite{espositoNCPA}). How to handle this issue while having an accurate OG estimation was discussed in a previous paper (\cite{chambou}). We remind here the reader of the main problem. The NCPA reference measurements are recorded around a diffraction-limited PSF and need to be re-scaled by the OG while working on sky: $s(\phi_{\text{NCPA}})\hookleftarrow s_{\phi}(\phi_{\text{NCPA}}) $. To compute $s_{\phi}(\phi_{\text{NCPA}})$ one needs to have an estimation of $T_{\phi}$ :

\begin{equation}
\begin{split} 
    s_{\phi}(\phi_{\text{NCPA}}) & = M_{\phi}.\phi_{\text{NCPA}}\\
    & = M_{\text{calib}}.T_{\phi}.\phi_{\text{NCPA}}
\end{split} 
\end{equation}

We expose here the results of a simulation where we use the GSC to handle NCPA in the AO loop. We keep the same simulations parameters as before (caption of Figure \ref{fig:fastOG}). The PyWFS modulation radius is $r_{mod}=3\  \lambda/D$ and $r0 = 14\ cm$. The interaction matrix is computed around a flat wavefront. We inject in our system 200 nm rms of NCPA distributed with a $f^{-2}$ power-law on the first 25 KL modes (except tip-tilt and focus). In that configuration and for a flat wavefront in the science path (H-band), the PSF in the wavefront sensing path (V-band) is given Figure \ref{fig:psfs_ncpa} (a) and the signal $\Omega_{\phi_{\text{NCPA}}}$ seen by the GSC is showed Figure \ref{fig:psfs_ncpa} (b).\\

We then proceed along the following way: we close the loop on the turbulence and after 5 seconds of closed loop operation, the NCPA are added in the system. These NCPA are then handled with different configurations and the results are compared with the NCPA-free case. Figure \ref{fig:ncpaSR} illustrates the results. The main conclusions from Figure \ref{fig:ncpaSR} are that:
\begin{enumerate}
    \item When one does not compensate for the NCPA (orange plot), the loop converges toward a flat wavefront in the sensing path, consequently inducing an important loss of SR in the science path, corresponding to the NCPA.
    \item When one uses a reference map $s(\phi_{\text{NCPA}})$ in the PyWFS measurement without updating it by the OG, it leads to a divergence of the loop (so-called NCPA catastrophe - yellow plot). This can be explained by the fact that because of the OG, the PyWFS introduces too much NCPA, creating an even stronger aberrated wavefront. This aberrated wavefront increases the OG on the next frame, which keep increasing the aberration, and so on. This is quickly making the loop to diverge.
    \item When one compensates the reference map by the time-averaged OG computed on the first 5 seconds of the loop thanks to a long exposure image of the GSC (purple plot), no NCPA catastrophe appears, and the final performance reaches an averaged SR of 82\%.
    \item When one compensates the reference map thanks to the OG computed at each frame, using the GSC camera (green plot), the final performance reaches an averaged SR of 86\%. This solution is better than the previous one because we keep track of the OG at each frame, and we also take in account the impact of the NCPA themselves on the OG. For illustration, the GSC image for a given closed-loop residual when compensating NCPA is given Figure \ref{fig:psfs_ncpa} (c).  
\end{enumerate}

\begin{figure}[!h]
    \centering
    \includegraphics[scale=0.45]{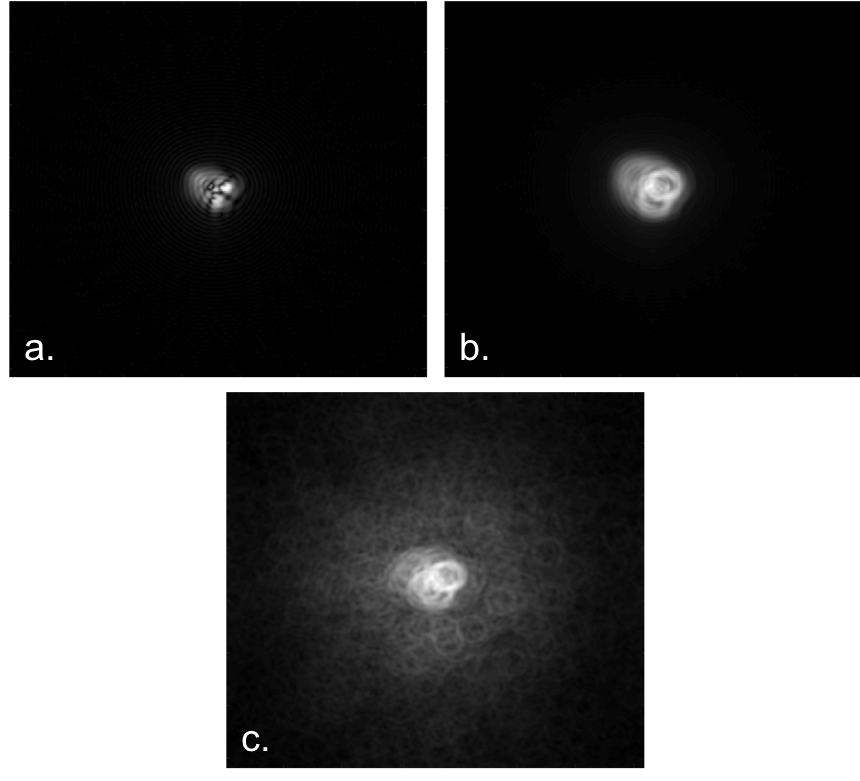}
    \caption{\textbf{a:} PSF on the pyramid apex when a flat wavefront is set in the science path. \textbf{b:} GSC signal when there are no residual phases and for a flat wavefront in the science path. \textbf{c:} GSC signal during closed-loop around NCPA.}
    \label{fig:psfs_ncpa}
\end{figure}

\begin{figure}[!h]
    \centering
    \includegraphics[scale=0.45]{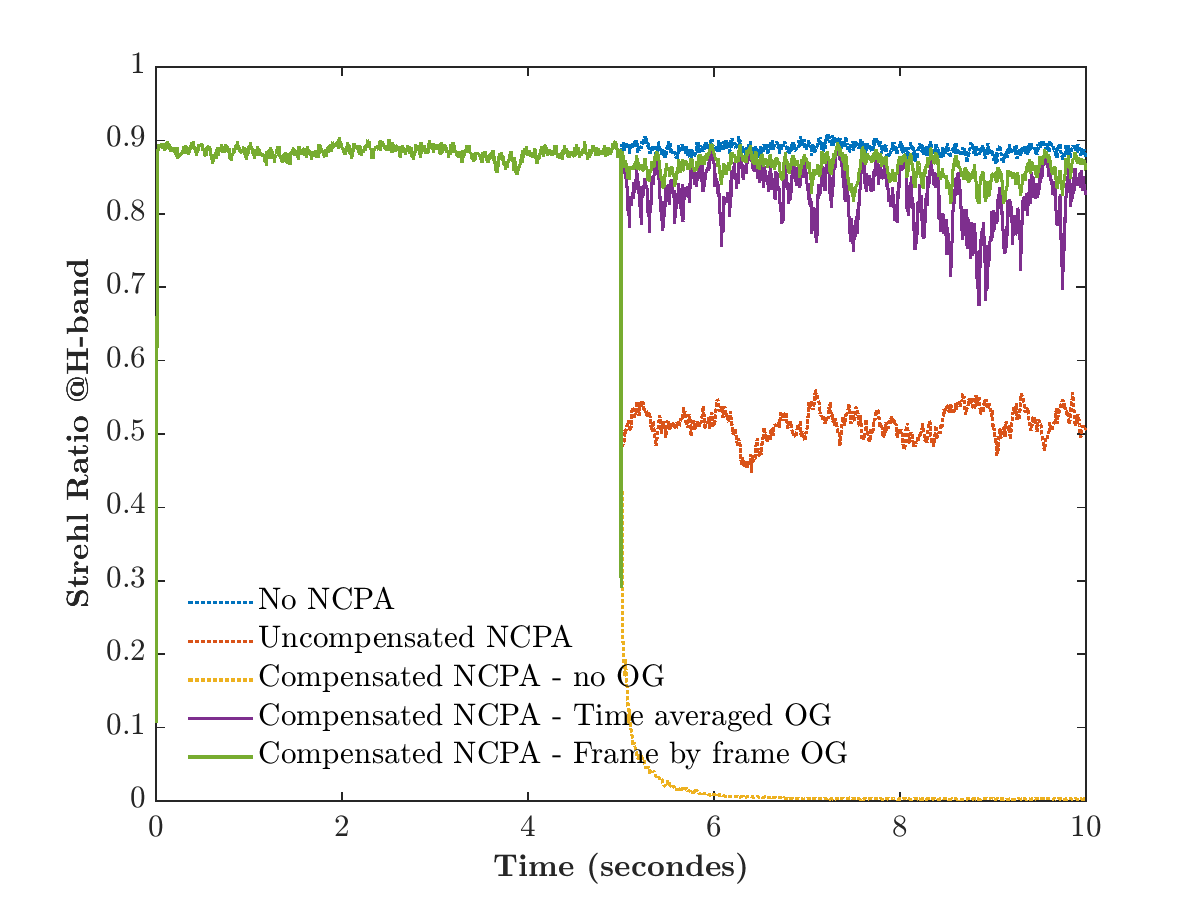}
    \caption{Strehl ratio for different cases of NCPA handling. In this simulation context, the case for which we compensate for NCPA without scaling by the OG lead to a diverging loop.}
    \label{fig:ncpaSR}
\end{figure}

This study is a clear demonstration that our strategy can solve AO control issue due to PyWFS OG. It also shows that even if the OG are compensated on a frame-by-frame basis, the ultimate performance (without NCPA) cannot be reached. This limitation is mainly due to the LPVS approach which is characterized by a linear description of the whole sensing problem. Improving the performance a step further would probably mean starting to look into other non-linear (2nd or 3rd order description) solutions which go beyond the simple matrix computation framework.

\section{Conclusion}

The PyWFS is a complex optical device exhibiting important non-linearities. One way to deal with this behaviour while keeping a matrix computation formalism is to consider the PyWFS as a Linear Parameter Varying System. To probe the sensing regime of this system at each measurement, one needs to implement a gain scheduling loop which gives an information on the sensor regime at every moment. With this perspective, the OG compensation can be deployed on a frame by frame basis. We provided here an innovative solution in that end: the Gain Scheduling Camera combined with a convolutive model.
As such, the PyWFS data synchronously merged, on a frame by frame basis, with a GSC data can be thought as a single WFS combining images from different light propagation planes. It therefore provides an efficient way to compensate for non-linearities at each AO loop frame without any delay, and significantly improve the final performance of the AO loop both in terms of sensitivity and dynamic range, as well as robustness. It also allows one to unambiguously disentangle the impact of OG from the full AO loop gain, which is a fundamental advantage for NCPA compensation.
The GSC solution has now to be implemented on the adaptive optics facility bench LOOPS at LAM for an experimental demonstration (\cite{LOOPS}).\\

\section{acknowledgements}
This work benefited from the support of the WOLF project ANR-18-CE31-0018 of the French National Research Agency (ANR). It has also been prepared as part of the activities of OPTICON H2020 (2017-2020) Work Package 1 (Calibration and test tools for AO assisted E-ELT instruments). OPTICON is supported by the Horizon 2020 Framework Programme of  the  European  Commission’s  (Grant  number  730890). Authors are acknowledging the support by the Action Spécifique Haute Résolution Angulaire (ASHRA) of CNRS/INSU co-funded by CNES. Vincent Chambouleyron PhD is co-funded by "Région Sud" and ONERA, in collaboration with First Light Imaging. Finally, part of this work is supported by the LabEx FOCUS ANR-11-LABX-0013.


\bibliographystyle{aa} 
\bibliography{aanda}
\end{document}